\documentclass[10pt,aps,prd,nofootinbib,superscriptaddress,
  twocolumn,preprintnumbers,balancelastpage]{revtex4}
\PassOptionsToPackage{slash}{url}

\usepackage{graphicx,epsfig,psfrag,amssymb,hyperref}
\usepackage{multirow}
\usepackage{color,graphicx,epsfig,psfrag,amsmath,empheq}
\usepackage{bm}
\usepackage{mathrsfs,amsfonts,color}
\usepackage{slashed}
\usepackage{hepunits}
\usepackage{color}
\usepackage{enumitem}
\usepackage{booktabs}
\usepackage{xspace}
\usepackage[utf8]{inputenc}

\AtBeginDocument{
\heavyrulewidth=.08em
\lightrulewidth=.05em
\cmidrulewidth=.03em
\belowrulesep=.65ex
\belowbottomsep=0pt
\aboverulesep=.4ex
\abovetopsep=0pt
\cmidrulesep=\doublerulesep
\cmidrulekern=.5em
\defaultaddspace=.5em
}

\long\def\symbolfootnote[#1]#2{\begingroup%
\def\thefootnote{\fnsymbol{footnote}}\footnote[#1]{#2}\endgroup}

\newcommand{\newc}{\newcommand}
\newc{\gsim}{\lower.7ex\hbox{$\;\stackrel{\textstyle>}{\sim}\;$}}
\newc{\lsim}{\lower.7ex\hbox{$\;\stackrel{\textstyle<}{\sim}\;$}}
\newc{\gev}{\,{\rm GeV}}
\newc{\mev}{\,{\rm MeV}}
\newc{\ev}{\,{\rm eV}}
\newc{\kev}{\,{\rm keV}}
\newc{\tev}{\,{\rm TeV}}
\newc{\MHT}{$H_T^{\text{miss}}$}
\newc{\MET}{$\slashed{E}_T$}
\newc{\MTT}{$M_{T2}$}

\newc{\mz}{M_Z}
\newc{\mpl}{M_*}
\newc{\mw}{m_{\rm weak}}
\newc{\nr}[1]{N^c_R{}_{#1}}

\def\beq{\begin{equation}}
\def\eeq{\end{equation}}
\newcommand{\bea}{\begin{eqnarray}\begin{aligned}}
\newcommand{\eea}{\end{aligned}\end{eqnarray}}
\def\bitem{\begin{itemize}}
\def\eitem{\end{itemize}}

\newcommand{\secvspace}{\vspace{5mm}}

\begin{document}

\title{DisCo Fever: Robust Networks Through Distance Correlation}


\author{Gregor Kasieczka}
\email{gregor.kasieczka@uni-hamburg.de}
\affiliation{ Institut f\"ur Experimentalphysik, Universit\"at Hamburg, 22761 Hamburg, Germany}

\author{David Shih}
\email{dshih@physics.rutgers.edu}
\affiliation{NHETC, Dept.~of Physics and Astronomy, Rutgers University, Piscataway, NJ 08854 USA}
\affiliation{Theory Group, Lawrence Berkeley National Laboratory, Berkeley, CA 94720, USA}
\affiliation{Berkeley Center for Theoretical Physics, University of California, Berkeley, CA 94720, USA}


\begin{abstract}
While deep learning has proven to be extremely successful at supervised classification tasks at the LHC and beyond, for practical applications, raw classification accuracy is often not the only consideration. One crucial issue is the stability of network predictions, either versus 
changes of individual features of the input data, or against systematic perturbations. We present a new 
method based on a novel application of ``distance correlation" (DisCo), a measure quantifying non-linear correlations,  that achieves equal performance to state-of-the-art adversarial decorrelation networks 
but is much simpler and more stable to train. To demonstrate the effectiveness of our method, we carefully recast a recent ATLAS study of decorrelation methods as applied to boosted, hadronic $W$-tagging. We also show the feasibility of DisCo regularization for more powerful convolutional neural networks, as well as for the problem of hadronic top tagging. 
\end{abstract}

\maketitle

\textbf{\textit{Introduction}} \newline
Recent breakthroughs in deep learning have begun to revolutionize many areas of high energy physics. One area that has received considerable focus is the problem of classifying different types of jets at the LHC. Deep neural networks have been applied, for example, to distinguishing top quarks from light quark and gluon jets. For this problem a large number of architectures based on fully connected neural networks~\cite{ANNTop,CanadaTop}, 
image-based methods~\cite{DeepTop,RuDeTop}, 
recursive clustering~\cite{RecursiveTop,CanadaLSTM}, 
physics variables~\cite{LoLa,LBN,NSubTop,LDA}, sets~\cite{EFN}, and graphs~\cite{ParticleNet,Moreno:2019neq} have been studied~\cite{CMSTop,ATLASTop,Landscape}. 
Related challenges of identifying vector bosons~\cite{DeepJetImages,Chen:2019uar}, b-quarks~\cite{ATL-PHYS-PUB-2017-003,DeepJet}, 
Higgs bosons~\cite{BenHiggs,Moreno:2019neq}, and distinguishing light quark from 
gluon jets~\cite{QGColor,QGReality,QGFC,Fraser:2018ieu} have seen similar progress. Beyond classifying single particles in an event, there is also work on developing holistic methods that classify full events according to the likely physics process that produced them~\cite{MartinTTH,Capsules}. Finally, some of these novel deep learning methods are beginning to be applied to concrete experimental analyses, see e.g.~\cite{Aaboud:2018wxv,Sirunyan:2019sza,Sirunyan:2019glc}.

So far, the recent activity in developing better jet classifiers with deep learning has focused on maximizing their  raw performance. 
However, the most accurate classifier is often not the best one for actual experimental applications. Instead, what is often desired is the most accurate classifier {\it given the constraint that it is decorrelated with one or more auxiliary variables}.

The underlying reason for this requirement is that classifiers are trained on Monte Carlo (MC) simulated examples (for which perfect truth labels are available), but are applied to (unlabeled) collision data. While the simulated events are of high fidelity, they do not perfectly reproduce the real data, and this gives rise to systematic differences between training and testing data. Understanding and mitigating these systematic differences is essential in any experimental analysis, and having a decorrelated classifier has many applications in this regard. For example, if the sources of systematic uncertainty are known, one can attempt to explicitly decorrelate a classifier against them in order to reduce or eliminate their effects~\cite{Louppe:2016ylz,UncertainSpanno,Windischhofer:2019ltt,UncertainQuast}. Or, one can attempt to control for these systematic differences using data-driven methods, such as sidebanding in the invariant mass.\footnote{
Although different auxiliary variables can be used in experimental analyses, one of the most common choices is invariant mass. So for concreteness, and without loss of generality, we will focus on the case of invariant mass for the remainder of this paper.} If the signal is localized but the background is smooth in mass, the sideband method allows one to calculate MC vs.\ data correction factors, define control samples, and estimate backgrounds. But if the classifier sculpts features (e.g.\ bumps) into the background mass distribution, it cannot be relied on for sidebanding. A classifier that is decorrelated with mass is sufficient (although not necessary) to guarantee smoothness of the background mass distribution. 

The issue is especially acute for powerful multivariate classifiers such as neural networks, which  will have a strong incentive to ``learn the mass" when building the optimal discriminant. Even if one excludes mass from the list of inputs to the machine learning algorithm, it may not be enough to achieve a decorrelated classifier -- many of the other inputs may be correlated with mass, and machine learning methods in general are flexible enough to exploit correlations of inputs. 
Such improvements will be especially relevant for (but not limited to) searches for new resonances with unknown mass. The identification of resonances in invariant mass distributions is historically the main avenue to discovery in experimental particle physics, and relies on robust background estimates. Therefore an important and significant challenge is to design classifiers that are as fully decorrelated from mass as possible while using maximal information.

In this paper we will present a new method for training decorrelated classifiers which achieves performance comparable to state-of-the-art methods, while being much easier to train. The key observation is that a statistical measure called {\it Distance Correlation} (DisCo) \cite{szekely2007, szekely2009, SzeKely:2013:DCT:2486206.2486394,szekely2014} is sensitive to general, nonlinear correlations between two random variables and can be efficiently computed from finite samples. 
Distance correlation is well-known in statistics and has been applied to various fields including data science~\cite{featureScreening} and biology~\cite{biostuff}. To our knowledge, this is the first application of DisCo to particle physics.

By including DisCo as an additive regularizer term in the loss function, we demonstrate that we can achieve a state-of-the art decorrelated classifier with just one additional hyperparameter (the coefficient of the DisCo regularizer). By varying this coefficient, we can control the tradeoff between classification performance and decorrelation, interpolating between a fully decorrelated tagger and a fully performant one.

To validate our methods and rigorously demonstrate that they are state-of-the-art, we will carefully reproduce the results of a recent ATLAS study of decorrelated  taggers for identifying boosted $W$ bosons \cite{ATL-PHYS-PUB-2018-014}. This study includes a comprehensive set of decorrelation methods, including \cite{DDT,CSS,Louppe:2016ylz,Shimmin:2017mfk}. The most promising technique so far (in terms of achieving the highest classifier performance for a given level of decorrelation) has been adversarially training a pair of neural networks: a classifier distinguishing different classes and an adversary predicting the mass~\cite{Louppe:2016ylz,Shimmin:2017mfk} for a given classifier output. 

The downside of the adversarial method has been that it is extremely difficult to implement in practice. Not only does one have to essentially train two separate neural networks, each with their own set of hyperparameters, but one has to carefully tune these two neural networks against each other. This stems from the nature of adversarial training: the objective is not to minimize a loss function, but rather to find a saddle point where the classifier loss is minimized but the adversary loss is maximized. Without careful tuning of learning rate schedules, number of epochs, minibatch sizes, etc., the training easily becomes unstable (since the loss is unbounded from below) and can quickly run away to a meaningless result. 

By contrast, DisCo regularization maintains the convex objective of the original loss function (i.e.\ the DisCo term is a positive measure of nonlinear correlations), making it much more stable to train. And since it only has one additional hyperparameter, no additional tuning is required. We will show, in the context of the ATLAS $W$-tagging study, that the result of DisCo decorrelation is comparable to that of adversarial decorrelation. 
In the Appendix, 
we will also demonstrate the state-of-the-art performance for top tagging with jet images and convolutional neural networks (CNNs).  

\secvspace
\secvspace

\textbf{\textit{Distance Correlation}} \newline
\label{disco}
Given a sample of paired vectors $(\vec x_i,\vec y_i)$ (where the index $i$ runs over the sample) drawn randomly from some distribution, we would like a function that measures the extent to which they are drawn from {\it independent} distributions, i.e.\ the extent to which $P_{joint}(\vec X,\vec Y)=P_X(\vec X)P_Y(\vec Y)$. In order for this function to be applicable in a deep learning context, we also require that this function be differentiable and that it can be  computed directly from the sample.

In our case the vectors are one dimensional and correspond to mass $X=m$  and classifier output $Y=y$ but clearly one can imagine many more applications of such a measure at the LHC and beyond. 

The usual Pearson correlation coefficient $R$ only measures linear dependencies so it is not suitable for our purposes.
Specifically, features can have nonlinear dependencies and still exhibit zero Pearson $R$.\footnote{Since the Pearson correlation coefficient is nonzero only if features are correlated, it can however be used to actively {\it correlate} features, see e.g.~\cite{DBLP:journals/corr/ChandarKLR15}.} There are many information-theoretic measures of similarity of distributions such as KL-divergence, Jensen-Shannon distance, and mutual information. These are difficult to compute directly from the sample, without binning. One can approximate these measures by training a classifier and using the likelihood ratio trick, but this again leads to adversarial methods, see e.g.~\cite{MINE, nowozin2016fgan,Windischhofer:2019ltt,mohamed2016learning,Cranmer:2015bka}.


One measure that seems to fit the bill perfectly is ``distance correlation", which  originated in the works of \cite{szekely2007, szekely2009, SzeKely:2013:DCT:2486206.2486394,szekely2014}. It can be computed from the sample and it has the key property that it is zero iff $X$ and $Y$ are independent. 

The definition of distance covariance is:
\beq\label{eq:dCov}
{\rm dCov}^2(X,Y)=\int d^p s d^q t\, |f_{X,Y}(s,t)-f_X(s)f_Y(t)|^2 w(s,t)
\eeq
where $X\in {\Bbb R}^p$, $Y\in {\Bbb R}^q$, $f_X$ and $f_Y$ are the characteristic functions for the random variables $X$ and $Y$, and $f_{X,Y}$ is the joint characteristic function for $X$ and $Y$. Finally 
\beq
w(s,t)\propto |s|^{-(p+1)}|t|^{-(q+1)}
\eeq
is a weight function that is uniquely determined up to an overall normalization by the requirement that ${\rm dCov}$ is invariant under constant shifts and orthogonal transformations, and equivariant under scale transformations \cite{SZEKELY20122278}. Since $f_{X,Y}=f_X f_Y$ iff $X$ and $Y$ are independent random variables, the definition (\ref{eq:dCov}) makes clear that distance covariance is a measure of the independence of $X$ and $Y$ that is zero iff $X$ and $Y$ are independent. 

Using the definition of the characteristic function it is straightforward to verify that we can also express ${\rm dCov}$ as
\bea\label{eq:dCovsample}
{\rm dCov}^2(X,Y) &= \langle |X-X'||Y-Y'|\rangle \cr
&\qquad + \langle |X-X'|\rangle\langle |Y-Y'|\rangle\cr
 &\qquad -2\langle |X-X'||Y-Y''|\rangle
\eea
where $|\cdot|$ refers to the Euclidean vector norm\footnote{In fact there is a family of distance covariance measures parameterized by $0<\alpha<2$ where one uses $|X-X'|^\alpha$ instead of $|X-X'|$. These relax the requirement of strict equivariance under rescalings. In this paper we will focus on $\alpha=1$ but in principle this would be another hyperparameter to explore.} and $(X,Y)$, $(X',Y')$, $(X'',Y'')$ are iid from the joint distribution of $(X,Y)$ ($X''$ is not used in (\ref{eq:dCovsample})). Using this alternative form of ${\rm dCov}^2$ it is straightforward to compute a sampling estimate of ${\rm dCov}^2$ from a dataset of $(x_i,y_i)$.\footnote{In the following we will be reweighting by $p_T$. So we actually need a {\it weighted} form of distance correlation. That follows easily from the sample definition (\ref{eq:dCovsample}).
}

Finally, we normalize the distance covariance by the individual distance variances to obtain {\it distance correlation}:
\beq
{\rm dCorr}^2(X,Y) = {{\rm dCov}^2(X,Y) \over {\rm dCov}(X,X) {\rm dCov}(Y,Y)}
\eeq
The distance correlation is bounded between 0 and 1. 
Normalizing ensures equally strong decorrelation independent of the overall scale. 

We will add ${\rm dCorr}^2$ as a regularizer term to the usual classifier loss function  in the following.\footnote{In principle another hyperparameter is the exact power of ${\rm dCorr}$ that one adds to the loss function. We have not explored this in much detail.} In detail:
\beq
L = L_{classifier}(\vec y,\vec y_{true}) + \lambda \, {\rm dCorr}_{y_{true}=0}^2(\vec m,\vec y)
\eeq
where $\lambda$ is a single hyperparameter that controls the tradeoff between classifier performance and decorrelation, $\vec y$ is the output of the NN on a single minibatch, and $\vec y_{true}$ and $\vec m$ are the true labels and masses respectively.\footnote{Our implementation of DisCo is available at \href{https://github.com/gkasieczka/DisCo/}{\texttt{https://github.com/gkasieczka/DisCo}}.} The subscript $y_{true}=0$ indicates that the distance correlation is only calculated for the subset of the minibatch that is background; this is the appropriate mode for $W$-tagging. Of course, for other applications it may be more appropriate to apply the decorrelation to all events, or even to signal events only.

\secvspace

\textbf{\textit{Samples}} \newline
\label{samples}
As discussed in the Introduction, we will focus in this paper on $W$ tagging, for which there is a detailed study of existing decorrelation methods by the ATLAS collaboration \cite{ATL-PHYS-PUB-2018-014}. (See the Appendix for a brief demonstration of DisCo decorrelation for top tagging.) By recasting the ATLAS study as closely as possible, we will be able to validate our methods and rigorously demonstrate that our method of distance correlation is state-of-the-art. 

Following the ATLAS study, we generate the SM processes $pp\to WW$ and $pp\to jj$ in {\sc Pythia~8.219} \cite{Sjostrand:2007gs} at $\sqrt{s}=13$~TeV with a generator level cut of $p_T>$250~GeV on the initial particles. We use  {\sc Delphes~3.4.1} with the default 
 detector card for detector simulation \cite{deFavereau:2013fsa}. We also use the built-in functionality of  {\sc Delphes} to simulate pileup with $\langle N_{PU}\rangle = 24$ as per the ATLAS study \cite{ATL-PHYS-PUB-2018-014}. 

\begin{figure}
\resizebox{1.\columnwidth}{!}{
\includegraphics[scale=1.0]{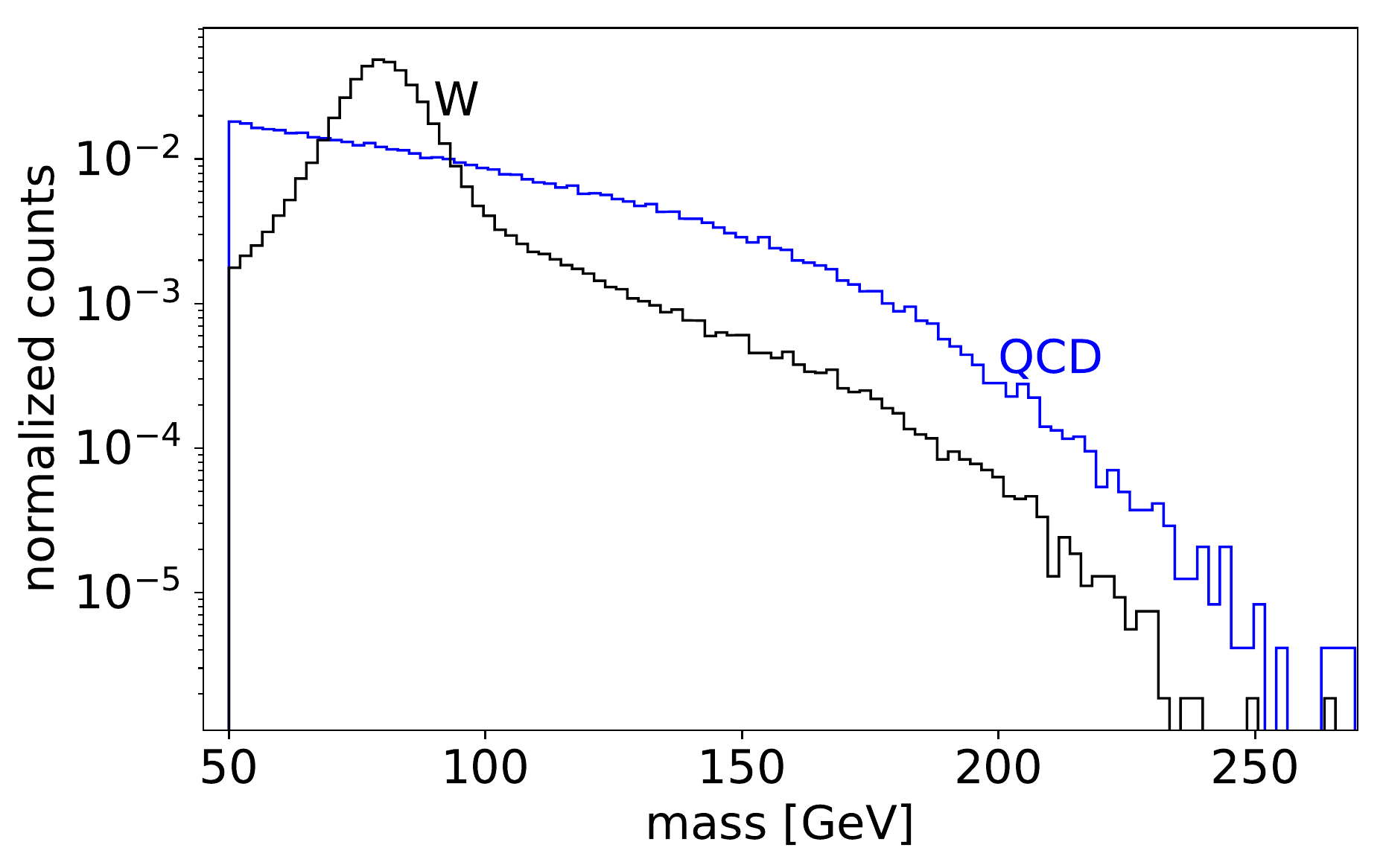}
}
\caption{Invariant mass distribution for the inclusive W and QCD samples.}
\label{fig:mass}
\end{figure}

Jets are reconstructed using 
\textsc{FastJet~3.0.1} \cite{Cacciari:2011ma} and the anti-$k_T$ algorithm \cite{Cacciari:2008gp} with $R=1$ distance parameter. Jets are required to have $|\eta|<2$ and to be within $\Delta R<0.75$ or the original parton. The daughters of the $W$ are also required to be within $\Delta R<0.75$ of the original $W$. Finally jets are trimmed \cite{Krohn:2009th} with parameters $R_{sub}=0.2$ and $f_{cut}=5$\%.  For the final sample, jets are required to have $m\in [50,300]$~GeV and $p_T\in[300,400]$~GeV; the mass  distributions for signal and background are shown in fig.~\ref{fig:mass}.
Apart from the very last requirement on $p_T$, these are all following the ATLAS study. Here we choose to focus on a more narrow range in $p_T$ for simplicity.

From this sample of jets, we compute the complete list of high-level kinematic variables shown in table 1 of the ATLAS study, see \cite{ATL-PHYS-PUB-2018-014} for more details and original references. These form the inputs for all the methods in the ATLAS study. We will also use them as inputs for the DNN plus distance correlation. 

\begin{figure}
\includegraphics[width=0.5\textwidth]{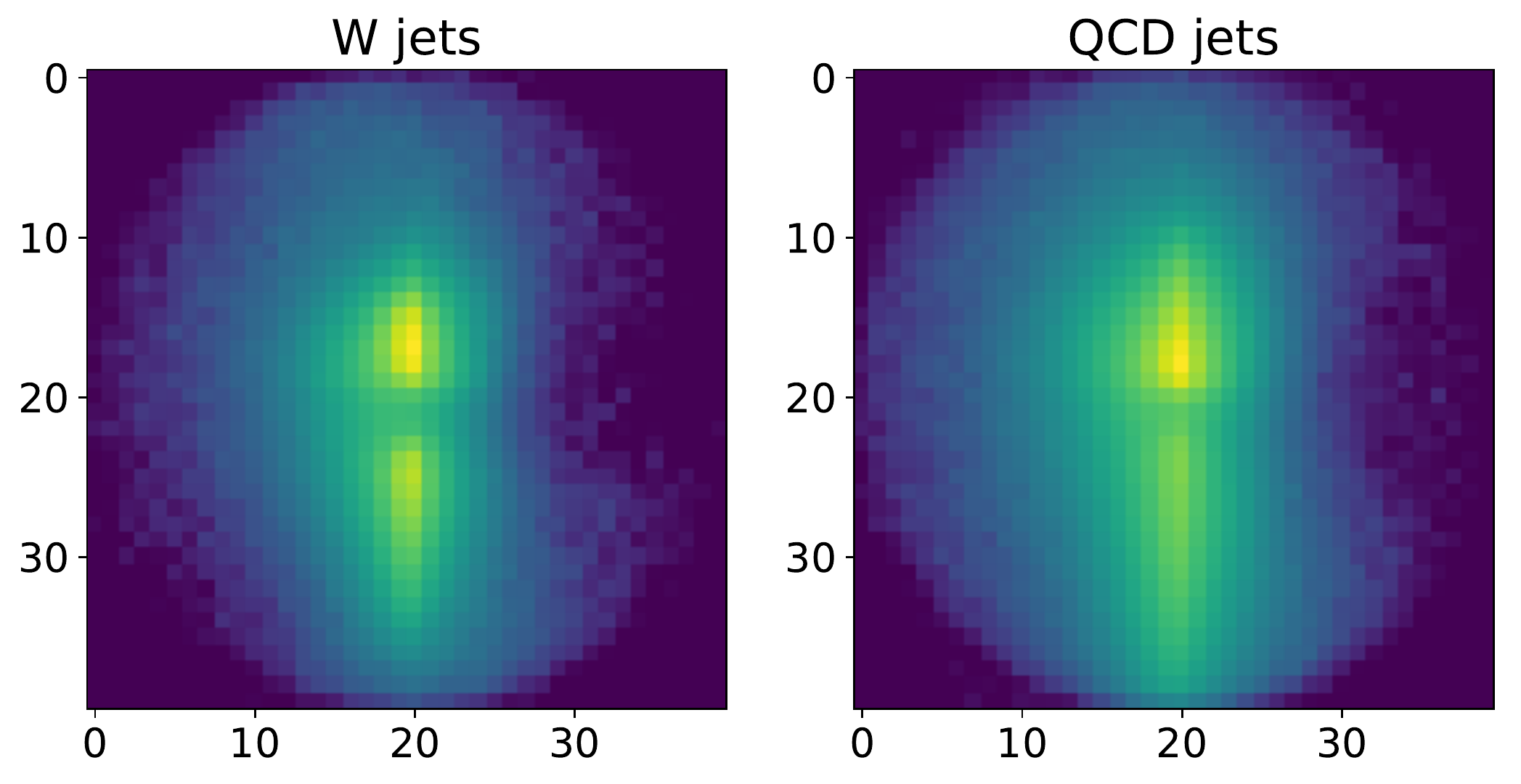}
\caption{Average of 100k jet images for W jets (left) and QCD jets (right). }
\label{fig:avgimg_W}
\end{figure}

Since we will also study the decorrelation of CNN classifiers (see below), we will also form jet images in the same way as \cite{Macaluso:2018tck}. We form images with $\Delta\eta=\Delta\phi=2$ and $40\times 40$ pixel resolution. For simplicity we stick to grayscale images (with pixel intensity equal to $p_T$) for this study. Fig.~\ref{fig:avgimg_W} shows the average of 100,000 $W$ and QCD jet images.

For all methods we reweight the training samples so that the $p_T$ distributions of signal and background are flat, following the ATLAS study. We use 50 evenly-spaced $p_T$ bins between 300 and 400 GeV. For evaluation ATLAS also reweights the signal $p_T$ distribution to look like background. But since we are taking such a narrow $p_T$ slice our $p_T$ distributions are basically identical, so we skip this step. 

All of the data samples used for this study will be made publicly available here \cite{gregor_kasieczka_2020_3606767}.

\secvspace

\textbf{\textit{Methods}} \newline
\label{methods}
Following~\cite{ATL-PHYS-PUB-2018-014} we measure the tagging performance by the rejection factor $R_{50}$ corresponding
to the inverse of the false positive rate (the probability to mis-identify a QCD jet as $W$ jet) at a true positive rate 
(the probability to correctly identify a $W$ jet) of $50\%$. The decorrelation is quantified by the inverse
of the Jensen-Shannon Divergence $1/\textrm{JSD}_{50}$ between the inclusive background distribution and the background distribution
passing the selection corresponding to a true positive rate of $50\%$.
The Jensen-Shannon Divergence is calculated from histograms with 50 bins between lowest and highest value. The binned entropy is measured in
 \textit{bits}.

We have implemented the following pairs of ($W$-tagging,\,decorrelation) methods in our work. From the ATLAS study: ($\tau_{21}$,\,DDT)~\cite{NSub,DDT}, 
($D_2$,\, kNN)~\cite{D21,D22,kNN}, (Adaboost BDT,\,uBoost)~\cite{uBoost}, and (DNN,\,adversary)~\cite{Louppe:2016ylz}. We will additionally include the simplest and possibly oldest decorrelation method, namely ``planing," or reweighting events so that the mass histograms of signal and background are identical. As this approach is relatively simple to implement and does not add much computational cost, it is a good baseline procedure.\footnote{See \cite{BryanOverview} for a recent comparison study of planing against other methods.} Finally, to all of this we will add our new method (DNN,\ DisCo regularization) for 
comparison. For details on all these methods, see the Appendix.

In addition, we will go beyond the ATLAS study and examine a CNN classifier acting on jet images, together with adversarial and DisCo decorrelation.  This will demonstrate that DisCo regularization is effective enough to decorrelate more powerful deep learning classifiers that use low-level, high-dimensional features. For the CNN classifier we use a scaled down version of the classifier in  \cite{Macaluso:2018tck}.  There are 4 convolutional layers with 64, 32, 32, 32 filters (size $4\times 4$), with $2\times 2$ Max pooling after the second and fourth layer. This is followed by 3 hidden layers with 32, 64 and 64 nodes. All activations are ReLU. Finally we output to softmax. 

For both CNN and DNN with DisCo regularization, we used the Adam optimizer with mini-batch size of 2048 and a fixed learning rate of $10^{-4}$. We found that the relatively large batch size of 2048 helped with the numerical stability of the DisCo regularizer.
We note 
that the sampling estimate (\ref{eq:dCovsample}) for distance covariance is known to be statistically biased, and an unbiased estimator was given in \cite{szekely2013partial}. The bias goes to zero as $\sim 1/n$ where $n$ is the size of the sample (the minibatch size in our case). We have verified that, as our minibatch size is sufficiently large, there is no practical benefit to using the unbiased estimate of distance covariance in our case.

For the DNN (CNN) we performed a scan in DisCo parameter $\lambda$ in the range 0--600 (0--250).  All classifiers were trained for 200 epochs; no early stopping was used. We have checked that 200 epochs is enough to ensure convergence, in the sense that training for more epochs does not improve things. Then, for each $\lambda$ and training instance, the model with the best validation loss is selected. This procedure is repeated six times with different random seeds to obtain a sense of the variability in the training outcomes. 

In all of the ML based methods we use 250k/80k/80k signal jets and 110k/330k/770k background jets for training/validation/testing. We use so many background jets in order to minimize the statistical error on the JSD calculation (which is calculated only for the background). 

The deep learning algorithms were implemented with PyTorch and trained on an NVIDIA P100 GPU.

\secvspace

\begin{figure}
\resizebox{1.\columnwidth}{!}{
\includegraphics[scale=1.0]{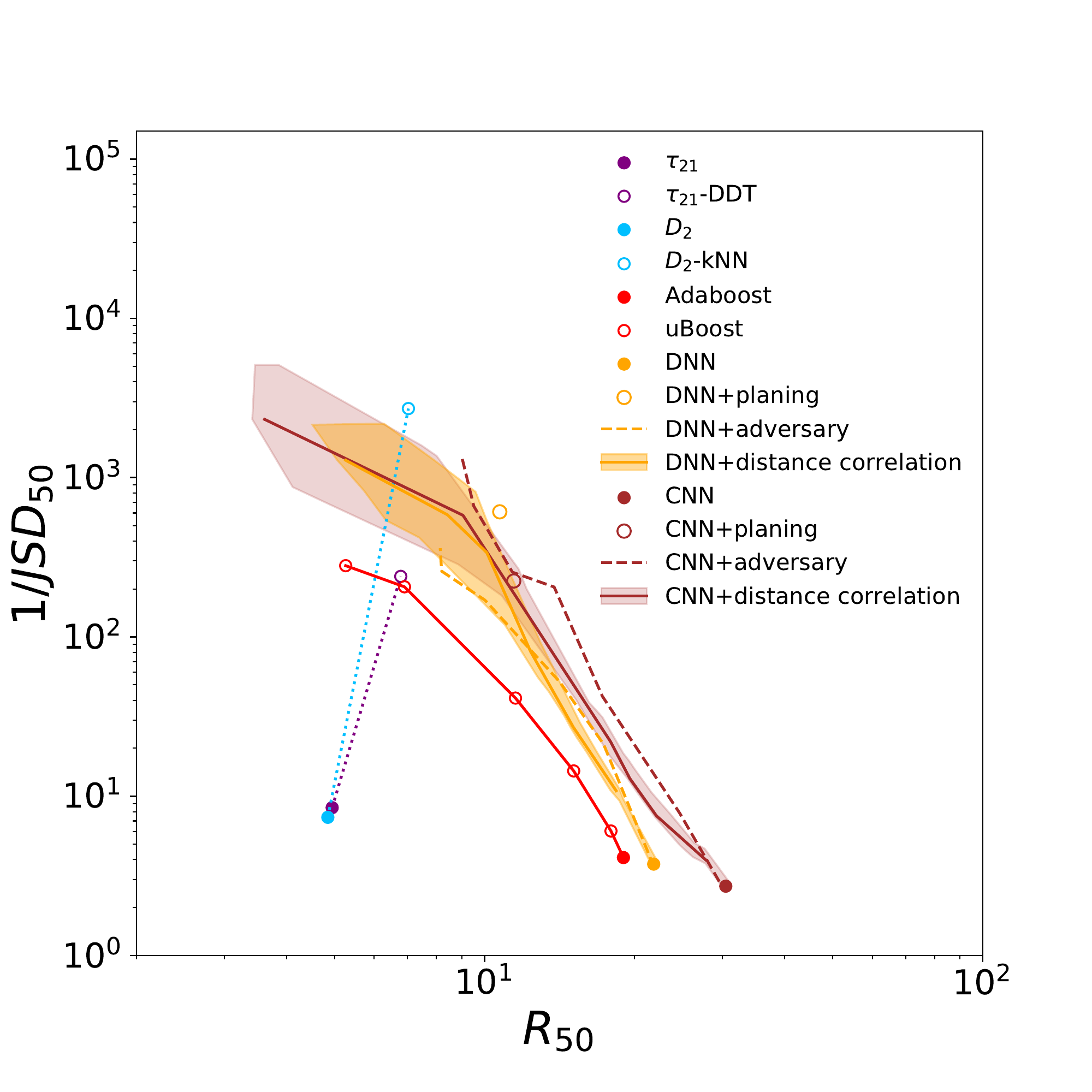}
}
\caption{{Decorrelation against background rejection for different approaches.}}
\label{fig:money}
\end{figure}

\textbf{\textit{Results}} \newline
\label{results}
Our final result is shown in fig.~\ref{fig:money}, where the performance of various decorrelation methods on the test set is summarized in the plane of $1/{\rm JSD}_{50}$ (which measures decorrelation)  vs.\ $R_{50}$ (which measures classifier performance). For DNN+DisCo and CNN+DisCo, the envelopes of the 6 independent trainings per $\lambda$ are shown, together with lines connecting the median-decorrelated points for different values of rejection. For the other ML methods, a representative result is shown per decorrelation parameter. (We have checked that the envelopes for DNN+adversary and CNN+adversary are comparable to their DisCo counterparts.)

The qualitative (and even quantitative) agreement with fig.~11(a) of~\cite{ATL-PHYS-PUB-2018-014}  is excellent, and we see a clear tradeoff between   classifier performance and the amount of decorrelation. 

\begin{figure}
\resizebox{\columnwidth}{!}{
\includegraphics[scale=0.8]{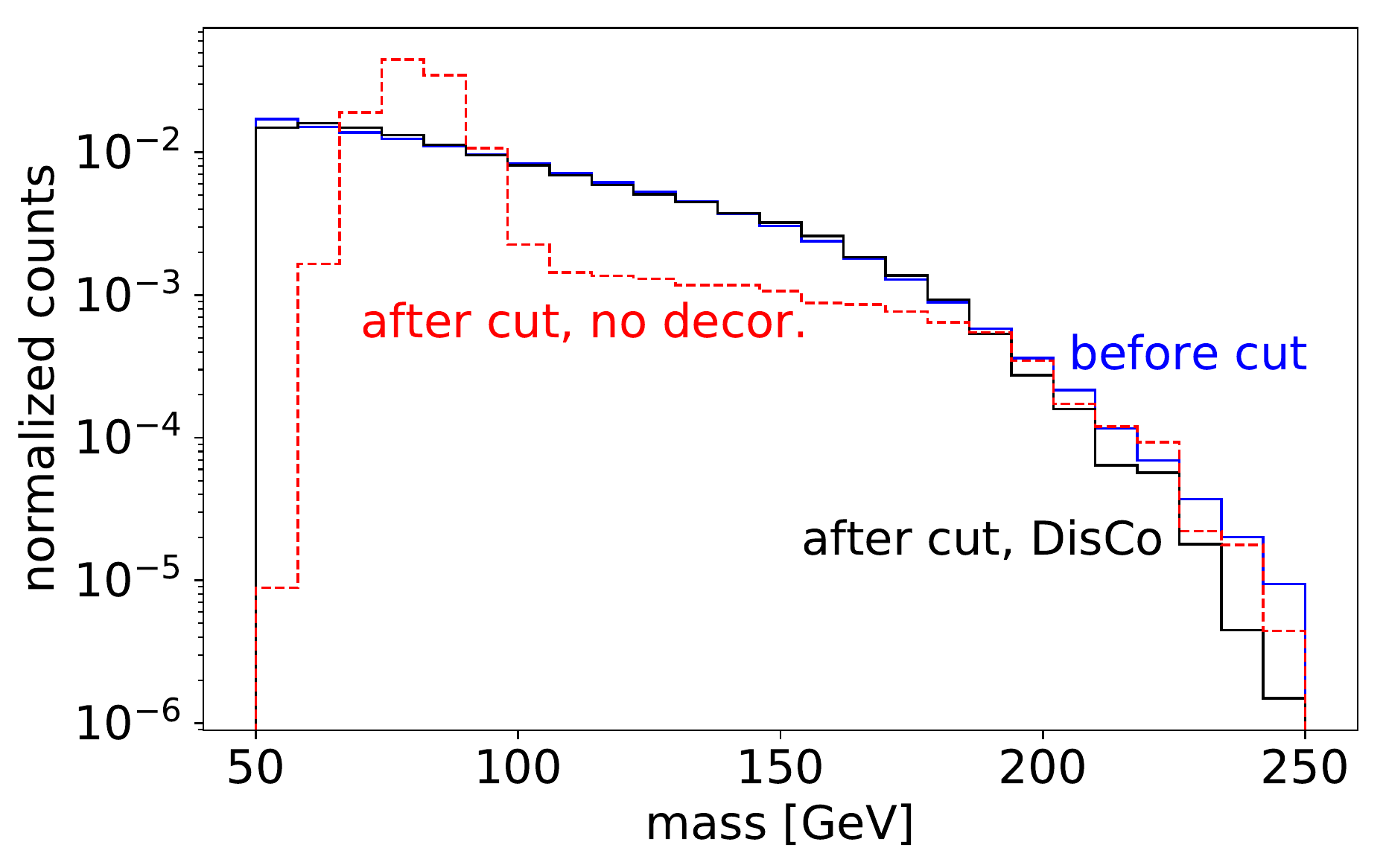}
}
\caption{QCD mass distribution before and after a cut on CNN plus DisCo ($W$-tagging) with signal efficiency of 50\% and ${\rm JSD}\sim 10^{-3}$.}
\label{fig:moneymass}
\end{figure}

Comparing DNN+DisCo to the other methods, we find that it has comparable performance to DNN+adversary. Meanwhile it is much easier to train -- whereas DisCo adds exactly one hyperparameter and no additional neural network parameters to the DNN, the adversary more than doubles the number of hyperparameters and adds an entire second NN to the story. See  the Appendix for a complete list of hyperparameters for the adversarial training. These were found through manual tuning and  their sheer complexity nicely illustrates the need for a simpler method of decorrelation.

We see that DisCo regularization is equally capable of decorrelating the more powerful CNN classifier, and again achieves comparable performance to CNN+adversary. One concern could have been that a more powerful deep learning method such as the CNN could overpower the DisCo regularizer, but our result demonstrates that this is not the case. At the highest levels of decorrelation, we note that both DNN and CNN performances are comparable.

In fig.~\ref{fig:moneymass}, we indicate more directly the level of decorrelation in the background mass distribution for the pure CNN case (no decorrelation), and for the CNN+DisCo method at a working point that achieves $1/{\rm JSD}_{50}\sim 10^{3}$.  We see that DisCo is quite effective at stabilizing the background mass distribution against a cut on the classifier. 

Finally, let us also comment briefly on the performance of planing. Unlike DisCo regularization and some of the other methods studied here, planing yields a single working point, instead of a tunable tradeoff between decorrelation and classifier performance. Since its performance depends on the joint probability distribution for mass and the other observables,\footnote{
Planing replaces $p(x,m)$  with $p(x,m)/p(m)$ which does not guarantee independence.} planing is not guaranteed to achieve strong results. But it is interesting to see that in this case (and in many of the cases studied in \cite{BryanOverview}), planing the DNN and CNN classifiers achieves very good performance. The performance lies on the DisCo regularization curve, and DisCo is capable of further decorrelation. 

\secvspace

\textbf{\textit{Conclusions}} \newline
\label{conclusions}
Deep learning is greatly increasing the classification performance for a wide number of reconstruction problems in particle physics. With the increasing adoption of these powerful machine learning solutions, a thorough understanding of their stability is needed. 

In this paper it was shown how a simple regularisation term based on the distance correlation metric can achieve state-of-the-art decorrelation power.  
Training is easier to set-up, with far less hyperparameters to optimise, and is more stable than adversarial networks, while simultaneously being more powerful than simpler approaches. 

DisCo regularization is an effective and promising new method for decorrelation which should have a host of immediate experimental applications at the LHC. At the same time, the potential use cases are much wider and include problems of fairness and bias of decision algorithms in social applications. This will be an extremely interesting direction for future exploration.

\begin{acknowledgments}

We thank Joern Bach, Ben Nachman, Bryan Ostdiek, Tilman Plehn, Matt Schwartz and Mike Williams  for helpful discussions. We are grateful to Chris Delitzsch, Steven Schramm and especially  Andreas Sogaard for help with details of the ATLAS decorrelation study. GK is supported by the Deutsche Forschungsgemeinschaft under Germany‘s Excellence Strategy – EXC 2121 ``Quantum Universe“ – 390833306". GK is grateful for the generous support and hospitality of the Rutgers NHETC Visitor Program where this work was initiated.  DS is supported by DOE grant DOE-SC0010008 and by the Director, Office of Science, Office of High Energy Physics of the U.S. Department of Energy under the Contract No. DE-AC02-05CH11231. DS thanks LBNL, BCTP and BCCP for their generous support and hospitality during his sabbatical year. 
\end{acknowledgments}


\appendix

\section{More details on the methods}
\label{appendix:methods}
\subsection{Planing}

One method to reduce correlation is to remove discriminating information carried by a variable.
The approach of giving weights to training events so the distributions for different classes are identical has been long used 
experimentally\footnote{See e.g.~\cite{Reweight2011,JME15002,ATLASRun1W,ATLASRun1Top} } and recently was studied for understanding network decisions~\cite{Planing} and resonance tagging~\cite{BryanOverview}. Specifically a weight 
$w_{i,C}$ for event with index $i$ of class $C$ is calculated by building a histogram of the feature $x$ so that $n_j$ denotes the number of events in bin $j$.\footnote{Due to the explicit use of histogramming, it can be difficult to generalise planing to multiple variables.} The weight can then be calculated as:
\begin{equation}
\left. w_{i,C} \right|_{x_i \textrm{ in bin } j} = A_C \frac{1}{n_j},
\end{equation}
where $A$ is a per-class normalisation factor.

Planing weights are then used in the training of an e.g. neural network classifier and modify the contribution of each event to the loss function. When applying the algorithm to events of unknown class in the testing phase no weights are used (i.e. weights are set equal to one).

\subsection{Designed decorrelated taggers}

For decorrelating a classifier for a single selection efficiency, a transformation of the output using the expected shape of the background distribution after the training is completed is possible as well~\cite{DDT}. This approach is named \textit{Designed decorrelated taggers} (\textit{DDT}). Concretely, to decorrelate feature $y$ against $x$, it is
transformed according to:
\begin{equation}
y' = y - M \cdot( x - O)
\end{equation}
where $O$ in an offset and $M$ is a slope parameter $\frac{dy}{dx}$ extracted for the background. 

\subsection{Fixed efficiency regression}

It is also possible to design decorrelated variables for non-linear relations between features by subtracting the expected
response for background examples~\cite{ATLASDecor}. This average response can also be parametrised against multiple features. Take for example the 
de-correlation of a feature $y$ against $x$ and $x'$. 

The decorrelated $y^{\textrm{k-NN}}$ can be calculated as
\begin{equation}
y^{\textrm{k-NN}} = y - y^{(P\;\%)}(x,x')
\end{equation}

with the threshold $y^{(P\;\%)}(x,x')$ corresponding to a true positive rate for background events $P$ interpoldated using a $k$-nearest neighbour regression fit~\cite{kNN}.

\subsection{uBoost}

The uBoost approach is a modified training methods for boosted decision trees (BDTs). A decision tree is a series of binary 
selection criteria that subsequently divide the data. Boosting refers to a combination of multiple decision trees to maximise a chosen 
classification metric such as the Gini coefficient or cross entropy. uBoost~\cite{uBoost} introduces an additional weight term in the boosting procedure so that regions in mass with low efficiency receive a higher weigth and regions with large efficiency receive a lower weight.

Following ATLAS, we used the implementation provided in the $\textsc{hep\_ml}$v0.6.0 package~\cite{hepml}. The hyperparameters were $\texttt{n\_estimators}=500$, $\texttt{learning\_rate}=0.5$, and $\texttt{base\_estimator}$ was the $\texttt{DecisionTreeClassifier}$ from $\texttt{sklearn}$ with $\texttt{max\_depth}=20$ and $\texttt{min\_samples\_leaf}=0.01$. For the uBoost uniforming rate (the analogue of $\lambda$ for DisCo and adversary), we scanned the range 0--3. We performed 5 independent trainings per uniforming rate and observed that the results were quite stable and consistent between them. Larger values of the uniforming rate were observed to populate lower $R_{50}$ but with a unreliably large variation in JSD$_{50}$, so they were not included in this study.

\subsection{DNN classifier}

As in the ATLAS study, we use for the DNN classifier a fully connected network consisting of 3 hidden layers with 64 nodes each. Except for the final softmax layer we use ReLU activations everywhere. Unlike the ATLAS study, we chose to include a batchnorm layer~\cite{BatchNorm} after the first hidden layer, as we found this improved the stability of the outcome. 

\subsection{Training with adversary}
Adversarial training follows the approach outlined in the Introduction, with the adversary
attempting to learn the PDF of the mass. The training objective is given by
\begin{equation}
\theta_C^* \theta_A^* = \textrm{arg}\;\textrm{min}_{\theta_C}\; \textrm{max}_{\theta_A}\; L_C - \lambda L_A
\end{equation}
and is usually solved by alternating training of the two networks. The different training objectives between discriminator and adversary 
are implemented using gradient reversal. Here $\lambda$ is a tunable hyperparameter defining the relative weight
of classification and decorrelation objective. The classifier loss term $L_C$ is the  usual cross-entropy term, while for the output of the adversary $A$ is the probability density produced by the Gaussian mixture model and $L_A = - \log A$ is evaluated at the true value of the mass.\footnote{An alternative approach to adversarial decorrelation attempts to infer the mass itself, in which case the adversarial loss $L_A(\theta_C, \theta_A)$ would take the form of a regression term or cross-entropy between different mass bins~\cite{Shimmin:2017mfk}.}

The adversary predicts a probability distribution function for the mass. The function is parametrised by a sum of 20 Gaussian 
distributions in a Gaussian mixture model. This means the network outputs 60 quantitites, interpreted as relative normalisation, mean and variance of 20 Gaussian distributions with a two layer fully connected network for each parameter where the first layer has 64 nodes and is shared. The output of the discriminator and $p_T$ are used as inputs to the adversary.

Training the adversary is done in three phases: only training the discriminator for 200 epochs (45 epochs for the CNN); only training the adversary with fixed discriminator for 20 epochs (30 epochs for the CNN); and joint training of both networks for 200 epochs (25 epochs for the CNN). For the DNN, 
the initial learning rates for the three phases are $\lambda_C=0.01$, $\lambda_A=0.05$  and $(\lambda_C,\lambda_A)=(3 \cdot 10^{-6}, 0.0001)$ respectively. The initial learning rates are subject to an exponential decay of $d_C=0.98$, $d_A=0.98$, $(d_C,d_A)=(0.97,0.97)$. For the CNN the initial learning rates are 
$\lambda_C=0.0001$, $\lambda_A=0.0005$,  and $(\lambda_C,\lambda_A)=(0.000001,0.001)$ for the three phases. No exponential decay is used for pre-training the classifier, and decay rates of $d_A=0.98$ and $(d_C,d_A)=(0.95,0.99)$ are used for the second and third phase. 
We verified for some representative values of $\lambda$ that greatly increasing the number of epochs (training up to 300 epochs) did not noticeably improve performance; nor did changing the model selection to the lowest loss instead of the final epoch.  The batch size is 8192 (1000) for the DNN (CNN) approach.

\section{Top tagging}
\label{appendix:toptagging}
\begin{figure}
\resizebox{1.0\columnwidth}{!}{
\vspace*{-3cm}
\includegraphics[scale=1.0]{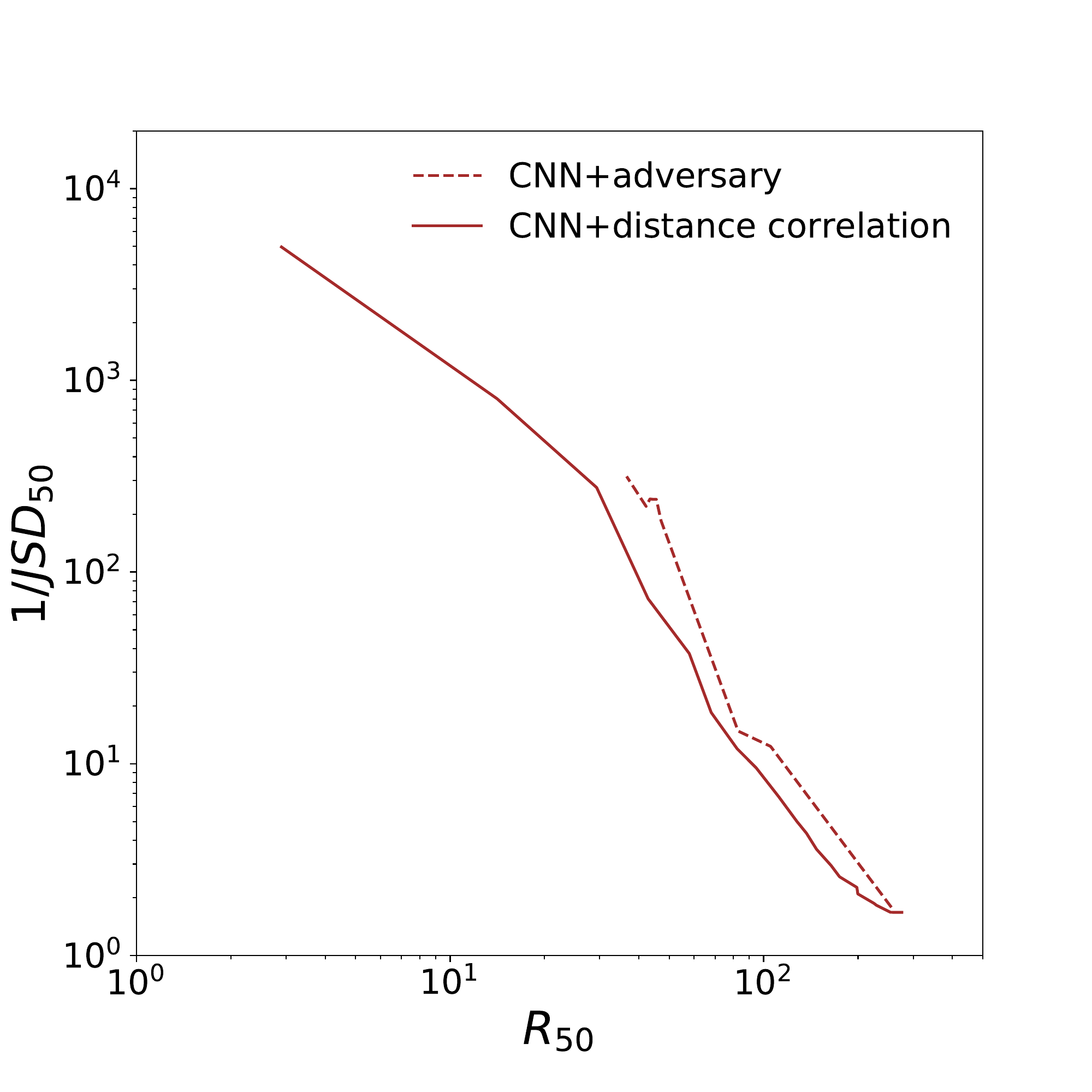}
}
\caption{Decorrelation against background rejection for adversarial decorrelation and DisCo in the case of top tagging.}
\vspace*{-0.5cm}
\label{fig:money_tops}
\end{figure}

We will also compare the performance of DisCo to adversarial decorrelation in the case of top tagging. 
For top tagging, we use the QCD and top samples in~\cite{Landscape}, and we restrict our comparison to CNNs trained on jet images (with the same specifications as the $W$-tagging). Fig.~\ref{fig:money_tops} shows the average top and QCD images. Despite the much higher possible discriminating power in top tagging, we again see that DisCo is comparable to the adversary, demonstrating that DisCo is indeed a powerful and sensitive measure of nonlinear correlation and a very effective penalty term for decorrelation.

\twocolumngrid
\vspace{-8pt}
\section*{References}
\vspace{-10pt}
\def\bibsection{}
\bibliographystyle{utphys}
\bibliography{refs}

\providecommand{\href}[2]{#2}\begingroup\raggedright\begin{thebibliography}{10}

\bibitem{ANNTop}
L.~G. Almeida, M.~Backović, M.~Cliche, S.~J. Lee, and M.~Perelstein,
  ``{Playing Tag with ANN: Boosted Top Identification with Pattern
  Recognition},'' \href{http://dx.doi.org/10.1007/JHEP07(2015)086}{{\em JHEP}
  {\bf 07} (2015)  086},
\href{http://arxiv.org/abs/1501.05968}{{\tt arXiv:1501.05968 [hep-ph]}}.

\bibitem{CanadaTop}
J.~Pearkes, W.~Fedorko, A.~Lister, and C.~Gay, ``{Jet Constituents for Deep
  Neural Network Based Top Quark Tagging},''
\href{http://arxiv.org/abs/1704.02124}{{\tt arXiv:1704.02124 [hep-ex]}}.

\bibitem{DeepTop}
G.~Kasieczka, T.~Plehn, M.~Russell, and T.~Schell, ``{Deep-learning Top Taggers
  or The End of QCD?},'' \href{http://dx.doi.org/10.1007/JHEP05(2017)006}{{\em
  JHEP} {\bf 05} (2017)  006},
\href{http://arxiv.org/abs/1701.08784}{{\tt arXiv:1701.08784 [hep-ph]}}.

\bibitem{RuDeTop}
S.~Macaluso and D.~Shih, ``{Pulling Out All the Tops with Computer Vision and
  Deep Learning},'' \href{http://dx.doi.org/10.1007/JHEP10(2018)121}{{\em JHEP}
  {\bf 10} (2018)  121},
\href{http://arxiv.org/abs/1803.00107}{{\tt arXiv:1803.00107 [hep-ph]}}.

\bibitem{RecursiveTop}
G.~Louppe, K.~Cho, C.~Becot, and K.~Cranmer, ``{QCD-Aware Recursive Neural
  Networks for Jet Physics},''
  \href{http://dx.doi.org/10.1007/JHEP01(2019)057}{{\em JHEP} {\bf 01} (2019)
  057},
\href{http://arxiv.org/abs/1702.00748}{{\tt arXiv:1702.00748 [hep-ph]}}.

\bibitem{CanadaLSTM}
S.~Egan, W.~Fedorko, A.~Lister, J.~Pearkes, and C.~Gay, ``{Long Short-Term
  Memory (LSTM) networks with jet constituents for boosted top tagging at the
  LHC},''
\href{http://arxiv.org/abs/1711.09059}{{\tt arXiv:1711.09059 [hep-ex]}}.

\bibitem{LoLa}
A.~Butter, G.~Kasieczka, T.~Plehn, and M.~Russell, ``{Deep-learned Top Tagging
  with a Lorentz Layer},''
  \href{http://dx.doi.org/10.21468/SciPostPhys.5.3.028}{{\em SciPost Phys.}
  {\bf 5} (2018) no.~3, 028},
\href{http://arxiv.org/abs/1707.08966}{{\tt arXiv:1707.08966 [hep-ph]}}.

\bibitem{LBN}
M.~Erdmann, E.~Geiser, Y.~Rath, and M.~Rieger, ``{Lorentz Boost Networks:
  Autonomous Physics-Inspired Feature Engineering},''
  \href{http://dx.doi.org/10.1088/1748-0221/14/06/P06006}{{\em JINST} {\bf 14}
  (2019) no.~06, P06006},
\href{http://arxiv.org/abs/1812.09722}{{\tt arXiv:1812.09722 [hep-ex]}}.

\bibitem{NSubTop}
L.~Moore, K.~Nordström, S.~Varma, and M.~Fairbairn, ``{Reports of My Demise
  Are Greatly Exaggerated: $N$-subjettiness Taggers Take On Jet Images},''
  \href{http://dx.doi.org/10.21468/SciPostPhys.7.3.036}{{\em SciPost Phys.}
  {\bf 7} (2019) no.~3, 036},
\href{http://arxiv.org/abs/1807.04769}{{\tt arXiv:1807.04769 [hep-ph]}}.

\bibitem{LDA}
B.~M. Dillon, D.~A. Faroughy, and J.~F. Kamenik, ``{Uncovering latent jet
  substructure},'' \href{http://dx.doi.org/10.1103/PhysRevD.100.056002}{{\em
  Phys. Rev.} {\bf D100} (2019) no.~5, 056002},
\href{http://arxiv.org/abs/1904.04200}{{\tt arXiv:1904.04200 [hep-ph]}}.

\bibitem{EFN}
P.~T. Komiske, E.~M. Metodiev, and J.~Thaler, ``{Energy Flow Networks: Deep
  Sets for Particle Jets},''
  \href{http://dx.doi.org/10.1007/JHEP01(2019)121}{{\em JHEP} {\bf 01} (2019)
  121},
\href{http://arxiv.org/abs/1810.05165}{{\tt arXiv:1810.05165 [hep-ph]}}.

\bibitem{ParticleNet}
H.~Qu and L.~Gouskos, ``{ParticleNet: Jet Tagging via Particle Clouds},''
\href{http://arxiv.org/abs/1902.08570}{{\tt arXiv:1902.08570 [hep-ph]}}.

\bibitem{Moreno:2019neq}
E.~A. Moreno, T.~Q. Nguyen, J.-R. Vlimant, O.~Cerri, H.~B. Newman, A.~Periwal,
  M.~Spiropulu, J.~M. Duarte, and M.~Pierini, ``{Interaction networks for the
  identification of boosted $H\to b\overline{b}$ decays},''
\href{http://arxiv.org/abs/1909.12285}{{\tt arXiv:1909.12285 [hep-ex]}}.

\bibitem{CMSTop}
{\bf CMS} Collaboration, ``{Machine learning-based identification of highly
  Lorentz-boosted hadronically decaying particles at the CMS experiment},''
{\em CMS-PAS-JME-18-002} (2019)  .

\bibitem{ATLASTop}
{\bf ATLAS} Collaboration, ``{Performance of top-quark and $W$-boson tagging
  with ATLAS in Run 2 of the LHC},''
  \href{http://dx.doi.org/10.1140/epjc/s10052-019-6847-8}{{\em Eur. Phys. J.}
  {\bf C79} (2019) no.~5, 375},
\href{http://arxiv.org/abs/1808.07858}{{\tt arXiv:1808.07858 [hep-ex]}}.

\bibitem{Landscape}
A.~Butter {\em et al.}, ``{The Machine Learning Landscape of Top Taggers},''
  \href{http://dx.doi.org/10.21468/SciPostPhys.7.1.014}{{\em SciPost Phys.}
  {\bf 7} (2019)  014},
\href{http://arxiv.org/abs/1902.09914}{{\tt arXiv:1902.09914 [hep-ph]}}.

\bibitem{DeepJetImages}
L.~de~Oliveira, M.~Kagan, L.~Mackey, B.~Nachman, and A.~Schwartzman,
  ``{Jet-images — deep learning edition},''
  \href{http://dx.doi.org/10.1007/JHEP07(2016)069}{{\em JHEP} {\bf 07} (2016)
  069},
\href{http://arxiv.org/abs/1511.05190}{{\tt arXiv:1511.05190 [hep-ph]}}.

\bibitem{Chen:2019uar}
Y.-C.~J. Chen, C.-W. Chiang, G.~Cottin, and D.~Shih, ``{Boosted $W/Z$ Tagging
  with Jet Charge and Deep Learning},''
\href{http://arxiv.org/abs/1908.08256}{{\tt arXiv:1908.08256 [hep-ph]}}.

\bibitem{ATL-PHYS-PUB-2017-003}
{\bf ATLAS} Collaboration, ``{Identification of Jets Containing $b$-Hadrons
  with Recurrent Neural Networks at the ATLAS Experiment},'' Tech. Rep.
  ATL-PHYS-PUB-2017-003, CERN, Geneva, Mar, 2017.
\newblock \url{http://cds.cern.ch/record/2255226}.

\bibitem{DeepJet}
{\bf CMS} Collaboration, ``{Performance of the DeepJet b tagging algorithm
  using 41.9/fb of data from proton-proton collisions at 13TeV with Phase 1 CMS
  detector},''{\em CMS-DP-2018-058} (Nov, 2018)  .
  \url{http://cds.cern.ch/record/2646773}.

\bibitem{BenHiggs}
J.~Lin, M.~Freytsis, I.~Moult, and B.~Nachman, ``{Boosting $H\to b\bar b$ with
  Machine Learning},'' \href{http://dx.doi.org/10.1007/JHEP10(2018)101}{{\em
  JHEP} {\bf 10} (2018)  101},
\href{http://arxiv.org/abs/1807.10768}{{\tt arXiv:1807.10768 [hep-ph]}}.

\bibitem{QGColor}
P.~T. Komiske, E.~M. Metodiev, and M.~D. Schwartz, ``{Deep learning in color:
  towards automated quark/gluon jet discrimination},''
  \href{http://dx.doi.org/10.1007/JHEP01(2017)110}{{\em JHEP} {\bf 01} (2017)
  110},
\href{http://arxiv.org/abs/1612.01551}{{\tt arXiv:1612.01551 [hep-ph]}}.

\bibitem{QGReality}
G.~Kasieczka, N.~Kiefer, T.~Plehn, and J.~M. Thompson, ``{Quark-Gluon Tagging:
  Machine Learning vs Detector},''
  \href{http://dx.doi.org/10.21468/SciPostPhys.6.6.069}{{\em SciPost Phys.}
  {\bf 6} (2019)  069},
\href{http://arxiv.org/abs/1812.09223}{{\tt arXiv:1812.09223 [hep-ph]}}.

\bibitem{QGFC}
H.~Luo, M.-x. Luo, K.~Wang, T.~Xu, and G.~Zhu, ``{Quark jet versus gluon jet:
  fully-connected neural networks with high-level features},''
  \href{http://dx.doi.org/10.1007/s11433-019-9390-8}{{\em Sci. China Phys.
  Mech. Astron.} {\bf 62} (2019) no.~9, 991011},
\href{http://arxiv.org/abs/1712.03634}{{\tt arXiv:1712.03634 [hep-ph]}}.

\bibitem{Fraser:2018ieu}
K.~Fraser and M.~D. Schwartz, ``{Jet Charge and Machine Learning},''
  \href{http://dx.doi.org/10.1007/JHEP10(2018)093}{{\em JHEP} {\bf 10} (2018)
  093},
\href{http://arxiv.org/abs/1803.08066}{{\tt arXiv:1803.08066 [hep-ph]}}.

\bibitem{MartinTTH}
M.~Erdmann, B.~Fischer, and M.~Rieger, ``{Jet-parton assignment in $t\bar t$H
  events using deep learning},''
  \href{http://dx.doi.org/10.1088/1748-0221/12/08/P08020}{{\em JINST} {\bf 12}
  (2017) no.~08, P08020},
\href{http://arxiv.org/abs/1706.01117}{{\tt arXiv:1706.01117 [hep-ex]}}.

\bibitem{Capsules}
S.~Diefenbacher, H.~Frost, G.~Kasieczka, T.~Plehn, and J.~M. Thompson,
  ``{CapsNets Continuing the Convolutional Quest},''
\href{http://arxiv.org/abs/1906.11265}{{\tt arXiv:1906.11265 [hep-ph]}}.

\bibitem{Aaboud:2018wxv}
{\bf ATLAS} Collaboration, ``{Search for pair production of heavy vector-like
  quarks decaying into hadronic final states in $pp$ collisions at $\sqrt{s} =
  13$ TeV with the ATLAS detector},''
  \href{http://dx.doi.org/10.1103/PhysRevD.98.092005}{{\em Phys. Rev.} {\bf
  D98} (2018) no.~9, 092005},
\href{http://arxiv.org/abs/1808.01771}{{\tt arXiv:1808.01771 [hep-ex]}}.

\bibitem{Sirunyan:2019sza}
{\bf CMS} Collaboration, ``{Search for pair production of vectorlike quarks in
  the fully hadronic final state},''
  \href{http://dx.doi.org/10.1103/PhysRevD.100.072001}{{\em Phys. Rev.} {\bf
  D100} (2019) no.~7, 072001},
\href{http://arxiv.org/abs/1906.11903}{{\tt arXiv:1906.11903 [hep-ex]}}.

\bibitem{Sirunyan:2019glc}
{\bf CMS} Collaboration, ``{Search for direct top squark pair production in
  events with one lepton, jets, and missing transverse momentum at 13 TeV with
  the CMS experiment},''  (2019)  ,
\href{http://arxiv.org/abs/1912.08887}{{\tt arXiv:1912.08887 [hep-ex]}}.

\bibitem{Louppe:2016ylz}
G.~Louppe, M.~Kagan, and K.~Cranmer, ``{Learning to Pivot with Adversarial
  Networks},''
\href{http://arxiv.org/abs/1611.01046}{{\tt arXiv:1611.01046 [stat.ME]}}.

\bibitem{UncertainSpanno}
C.~Englert, P.~Galler, P.~Harris, and M.~Spannowsky, ``{Machine Learning
  Uncertainties with Adversarial Neural Networks},''
  \href{http://dx.doi.org/10.1140/epjc/s10052-018-6511-8}{{\em Eur. Phys. J.}
  {\bf C79} (2019) no.~1, 4},
\href{http://arxiv.org/abs/1807.08763}{{\tt arXiv:1807.08763 [hep-ph]}}.

\bibitem{Windischhofer:2019ltt}
P.~Windischhofer, M.~Zgubič, and D.~Bortoletto, ``{Preserving physically
  important variables in optimal event selections: A case study in Higgs
  physics},''
\href{http://arxiv.org/abs/1907.02098}{{\tt arXiv:1907.02098 [hep-ph]}}.

\bibitem{UncertainQuast}
S.~Wunsch, S.~Jörger, R.~Wolf, and G.~Quast, ``{Reducing the dependence of the
  neural network function to systematic uncertainties in the input space},''
\href{http://arxiv.org/abs/1907.11674}{{\tt arXiv:1907.11674
  [physics.data-an]}}.

\bibitem{szekely2007}
G.~J. Székely, M.~L. Rizzo, and N.~K. Bakirov, ``Measuring and testing
  dependence by correlation of distances,''
  \href{http://dx.doi.org/10.1214/009053607000000505}{{\em Ann. Statist.} {\bf
  35} (2007) no.~6, 2769--2794}.
  \url{https://doi.org/10.1214/009053607000000505}.

\bibitem{szekely2009}
G.~J. Székely and M.~L. Rizzo, ``Brownian distance covariance,''
  \href{http://dx.doi.org/10.1214/09-AOAS312}{{\em Ann. Appl. Stat.} {\bf 3}
  (2009) no.~4, 1236--1265}. \url{https://doi.org/10.1214/09-AOAS312}.

\bibitem{SzeKely:2013:DCT:2486206.2486394}
G.~J. Sz{\'e}kely and M.~L. Rizzo, ``The distance correlation t-test of
  independence in high dimension,''
  \href{http://dx.doi.org/10.1016/j.jmva.2013.02.012}{{\em J. Multivar. Anal.}
  {\bf 117} (2013)  193--213}.
  \url{http://dx.doi.org/10.1016/j.jmva.2013.02.012}.

\bibitem{szekely2014}
G.~J. Székely and M.~L. Rizzo, ``Partial distance correlation with methods for
  dissimilarities,'' \href{http://dx.doi.org/10.1214/14-AOS1255}{{\em Ann.
  Statist.} {\bf 42} (2014) no.~6, 2382--2412}.
  \url{https://doi.org/10.1214/14-AOS1255}.

\bibitem{featureScreening}
R.~Li, W.~Zhong, and L.~Zhu, ``Feature screening via distance correlation
  learning,'' \href{http://dx.doi.org/10.1080/01621459.2012.695654}{{\em
  Journal of the American Statistical Association} {\bf 107} (2012) no.~499,
  1129--1139}.

\bibitem{biostuff}
A.~Villaverde and J.~Banga,
  \href{http://dx.doi.org/10.1098/rsif.2013.0505}{``Reverse engineering and
  identification in systems biology: Strategies, perspectives and
  challenges,''{\em Journal of the Royal Society} {\bf 11} (02, 2014)
  20130505}.

\bibitem{ATL-PHYS-PUB-2018-014}
{\bf ATLAS} Collaboration, ``{Performance of mass-decorrelated jet substructure
  observables for hadronic two-body decay tagging in ATLAS},''{\em
  ATL-PHYS-PUB-2018-014} (Jul, 2018)  .
  \url{http://cds.cern.ch/record/2630973}.

\bibitem{DDT}
J.~Dolen, P.~Harris, S.~Marzani, S.~Rappoccio, and N.~Tran, ``{Thinking outside
  the ROCs: Designing Decorrelated Taggers (DDT) for jet substructure},''
  \href{http://dx.doi.org/10.1007/JHEP05(2016)156}{{\em JHEP} {\bf 05} (2016)
  156},
\href{http://arxiv.org/abs/1603.00027}{{\tt arXiv:1603.00027 [hep-ph]}}.

\bibitem{CSS}
I.~Moult, B.~Nachman, and D.~Neill, ``{Convolved Substructure: Analytically
  Decorrelating Jet Substructure Observables},''
  \href{http://dx.doi.org/10.1007/JHEP05(2018)002}{{\em JHEP} {\bf 05} (2018)
  002},
\href{http://arxiv.org/abs/1710.06859}{{\tt arXiv:1710.06859 [hep-ph]}}.

\bibitem{Shimmin:2017mfk}
C.~Shimmin, P.~Sadowski, P.~Baldi, E.~Weik, D.~Whiteson, E.~Goul, and
  A.~Søgaard, ``{Decorrelated Jet Substructure Tagging using Adversarial
  Neural Networks},'' \href{http://dx.doi.org/10.1103/PhysRevD.96.074034}{{\em
  Phys. Rev.} {\bf D96} (2017) no.~7, 074034},
\href{http://arxiv.org/abs/1703.03507}{{\tt arXiv:1703.03507 [hep-ex]}}.

\bibitem{DBLP:journals/corr/ChandarKLR15}
S.~Chandar, M.~M. Khapra, H.~Larochelle, and B.~Ravindran, ``Correlational
  neural networks,'' {\em CoRR} {\bf abs/1504.07225} (2015)  ,
  \href{http://arxiv.org/abs/1504.07225}{{\tt arXiv:1504.07225}}.
  \url{http://arxiv.org/abs/1504.07225}.

\bibitem{MINE}
M.~I. Belghazi, A.~Baratin, S.~Rajeswar, S.~Ozair, Y.~Bengio, A.~Courville, and
  R.~D. Hjelm, ``Mine: Mutual information neural estimation,''
  \href{http://arxiv.org/abs/1801.04062}{{\tt arXiv:1801.04062 [cs.LG]}}.

\bibitem{nowozin2016fgan}
S.~Nowozin, B.~Cseke, and R.~Tomioka, ``f-gan: Training generative neural
  samplers using variational divergence minimization,'' 2016.

\bibitem{mohamed2016learning}
S.~Mohamed and B.~Lakshminarayanan, ``Learning in implicit generative models,''
  \href{http://arxiv.org/abs/1610.03483}{{\tt arXiv:1610.03483 [stat.ML]}}.

\bibitem{Cranmer:2015bka}
K.~Cranmer, J.~Pavez, and G.~Louppe, ``{Approximating Likelihood Ratios with
  Calibrated Discriminative Classifiers},''
  \href{http://arxiv.org/abs/1506.02169}{{\tt arXiv:1506.02169 [stat.AP]}}.

\bibitem{SZEKELY20122278}
G.~J. Székely and M.~L. Rizzo, ``On the uniqueness of distance covariance,''
  \href{http://dx.doi.org/https://doi.org/10.1016/j.spl.2012.08.007}{{\em
  Statistics \& Probability Letters} {\bf 82} (2012) no.~12, 2278 -- 2282}.
  \url{http://www.sciencedirect.com/science/article/pii/S0167715212003124}.

\bibitem{Sjostrand:2007gs}
T.~Sjostrand, S.~Mrenna, and P.~Z. Skands, ``{A Brief Introduction to PYTHIA
  8.1},'' \href{http://dx.doi.org/10.1016/j.cpc.2008.01.036}{{\em Comput. Phys.
  Commun.} {\bf 178} (2008)  852--867},
\href{http://arxiv.org/abs/0710.3820}{{\tt arXiv:0710.3820 [hep-ph]}}.

\bibitem{deFavereau:2013fsa}
{\bf DELPHES 3} Collaboration, J.~de~Favereau, C.~Delaere, P.~Demin,
  A.~Giammanco, V.~Lemaître, A.~Mertens, and M.~Selvaggi, ``{DELPHES 3, A
  modular framework for fast simulation of a generic collider experiment},''
  \href{http://dx.doi.org/10.1007/JHEP02(2014)057}{{\em JHEP} {\bf 02} (2014)
  057},
\href{http://arxiv.org/abs/1307.6346}{{\tt arXiv:1307.6346 [hep-ex]}}.

\bibitem{Cacciari:2011ma}
M.~Cacciari, G.~P. Salam, and G.~Soyez, ``{FastJet User Manual},''
  \href{http://dx.doi.org/10.1140/epjc/s10052-012-1896-2}{{\em Eur. Phys. J.}
  {\bf C72} (2012)  1896},
\href{http://arxiv.org/abs/1111.6097}{{\tt arXiv:1111.6097 [hep-ph]}}.

\bibitem{Cacciari:2008gp}
M.~Cacciari, G.~P. Salam, and G.~Soyez, ``{The anti-$k_t$ jet clustering
  algorithm},'' \href{http://dx.doi.org/10.1088/1126-6708/2008/04/063}{{\em
  JHEP} {\bf 04} (2008)  063},
\href{http://arxiv.org/abs/0802.1189}{{\tt arXiv:0802.1189 [hep-ph]}}.

\bibitem{Krohn:2009th}
D.~Krohn, J.~Thaler, and L.-T. Wang, ``{Jet Trimming},''
  \href{http://dx.doi.org/10.1007/JHEP02(2010)084}{{\em JHEP} {\bf 02} (2010)
  084},
\href{http://arxiv.org/abs/0912.1342}{{\tt arXiv:0912.1342 [hep-ph]}}.

\bibitem{Macaluso:2018tck}
S.~Macaluso and D.~Shih, ``{Pulling Out All the Tops with Computer Vision and
  Deep Learning},'' \href{http://dx.doi.org/10.1007/JHEP10(2018)121}{{\em JHEP}
  {\bf 10} (2018)  121},
\href{http://arxiv.org/abs/1803.00107}{{\tt arXiv:1803.00107 [hep-ph]}}.

\bibitem{gregor_kasieczka_2020_3606767}
G.~Kasieczka and D.~Shih, ``Datasets for boosted $w$ tagging,'' Jan., 2020.
\newblock \url{https://doi.org/10.5281/zenodo.3606767}.

\bibitem{NSub}
J.~Thaler and K.~Van~Tilburg, ``{Identifying Boosted Objects with
  N-subjettiness},'' \href{http://dx.doi.org/10.1007/JHEP03(2011)015}{{\em
  JHEP} {\bf 03} (2011)  015},
\href{http://arxiv.org/abs/1011.2268}{{\tt arXiv:1011.2268 [hep-ph]}}.

\bibitem{D21}
A.~J. Larkoski, I.~Moult, and D.~Neill, ``{Power Counting to Better Jet
  Observables},'' \href{http://dx.doi.org/10.1007/JHEP12(2014)009}{{\em JHEP}
  {\bf 12} (2014)  009},
\href{http://arxiv.org/abs/1409.6298}{{\tt arXiv:1409.6298 [hep-ph]}}.

\bibitem{D22}
A.~J. Larkoski, I.~Moult, and D.~Neill, ``{Analytic Boosted Boson
  Discrimination},'' \href{http://dx.doi.org/10.1007/JHEP05(2016)117}{{\em
  JHEP} {\bf 05} (2016)  117},
\href{http://arxiv.org/abs/1507.03018}{{\tt arXiv:1507.03018 [hep-ph]}}.

\bibitem{kNN}
S.~A. Dudani, ``The distance-weighted k-nearest-neighbor rule,''
  \href{http://dx.doi.org/10.1109/tsmc.1976.5408784}{{\em {IEEE} Transactions
  on Systems, Man, and Cybernetics} {\bf {SMC}-6} (1976) no.~4, 325--327}.
  \url{https://doi.org/10.1109%2Ftsmc.1976.5408784}.

\bibitem{uBoost}
J.~Stevens and M.~Williams, ``{uBoost: A boosting method for producing uniform
  selection efficiencies from multivariate classifiers},''
  \href{http://dx.doi.org/10.1088/1748-0221/8/12/P12013}{{\em JINST} {\bf 8}
  (2013)  P12013},
\href{http://arxiv.org/abs/1305.7248}{{\tt arXiv:1305.7248 [nucl-ex]}}.

\bibitem{BryanOverview}
L.~Bradshaw, R.~K. Mishra, A.~Mitridate, and B.~Ostdiek, ``{Mass Agnostic Jet
  Taggers},''
\href{http://arxiv.org/abs/1908.08959}{{\tt arXiv:1908.08959 [hep-ph]}}.

\bibitem{szekely2013partial}
G.~J. Szekely and M.~L. Rizzo, ``Partial distance correlation with methods for
  dissimilarities,'' \href{http://arxiv.org/abs/1310.2926}{{\tt arXiv:1310.2926
  [stat.ME]}}.

\bibitem{Reweight2011}
J.~Freeman, J.~D. Lewis, W.~Ketchum, S.~Poprocki, A.~Pronko, V.~Rusu, and
  P.~Wittich, ``{An Artificial neural network based $b$ jet identification
  algorithm at the CDF Experiment},''
  \href{http://dx.doi.org/10.1016/j.nima.2011.10.024}{{\em Nucl. Instrum.
  Meth.} {\bf A663} (2012)  37--47},
\href{http://arxiv.org/abs/1108.4738}{{\tt arXiv:1108.4738 [hep-ex]}}.

\bibitem{JME15002}
{\bf CMS} Collaboration, ``{Top Tagging with New Approaches},''
{\em CMS-PAS-JME-15-002} (2016)  .

\bibitem{ATLASRun1W}
{\bf ATLAS} Collaboration, ``{Identification of boosted, hadronically decaying
  W bosons and comparisons with ATLAS data taken at $\sqrt{s} = 8$ TeV},''
  \href{http://dx.doi.org/10.1140/epjc/s10052-016-3978-z}{{\em Eur. Phys. J.}
  {\bf C76} (2016) no.~3, 154},
\href{http://arxiv.org/abs/1510.05821}{{\tt arXiv:1510.05821 [hep-ex]}}.

\bibitem{ATLASRun1Top}
{\bf ATLAS} Collaboration, ``{Identification of high transverse momentum top
  quarks in $pp$ collisions at $\sqrt{s}$ = 8 TeV with the ATLAS detector},''
  \href{http://dx.doi.org/10.1007/JHEP06(2016)093}{{\em JHEP} {\bf 06} (2016)
  093},
\href{http://arxiv.org/abs/1603.03127}{{\tt arXiv:1603.03127 [hep-ex]}}.

\bibitem{Planing}
S.~Chang, T.~Cohen, and B.~Ostdiek, ``{What is the Machine Learning?},''
  \href{http://dx.doi.org/10.1103/PhysRevD.97.056009}{{\em Phys. Rev.} {\bf
  D97} (2018) no.~5, 056009},
\href{http://arxiv.org/abs/1709.10106}{{\tt arXiv:1709.10106 [hep-ph]}}.

\bibitem{ATLASDecor}
{\bf ATLAS Collaboration} Collaboration, ``{Performance of mass-decorrelated
  jet substructure observables for hadronic two-body decay tagging in ATLAS},''
  Tech. Rep. ATL-PHYS-PUB-2018-014, CERN, Geneva, Jul, 2018.
\newblock \url{http://cds.cern.ch/record/2630973}.

\bibitem{hepml}
A.~Rogozhnikov {\em et al.}, ``{hep\_ml: Machine Learning for High Energy
  Physics, version 0.6},''. \url{https://github.com/arogozhnikov/hep_ml}.

\bibitem{BatchNorm}
S.~Ioffe and C.~Szegedy, ``Batch normalization: Accelerating deep network
  training by reducing internal covariate shift,''
  \href{http://arxiv.org/abs/1502.03167}{{\tt arXiv:1502.03167 [cs.LG]}}.

\end{thebibliography}\endgroup

\end{document}